\documentclass{article}
\usepackage{spconf,amsmath,graphicx}
\usepackage{amsmath,amsfonts,amssymb}
\usepackage[nolist, nohyperlinks]{acronym}
\usepackage[table]{xcolor}  %
\usepackage{booktabs,makecell,multirow}
\usepackage{xspace} %
\usepackage{flushend}
\usepackage[table]{xcolor} %
\usepackage{pgf}          %
\usepackage{bm}

\usepackage{tikzpagenodes} %
\usepackage{tikz}
\usepackage{atbegshi}

\usepackage{microtype}
\microtypesetup{protrusion=true, expansion=true}

\usepackage{hyperref}

\def\x{{\mathbf x}}

\title{Are Modern Speech Enhancement Systems Vulnerable to Adversarial Attacks?}
\name{Rostislav Makarov$^1$, Lea Schönherr$^{2}$, Timo Gerkmann$^1$}
\address{$^1$ Signal Processing (SP), University of Hamburg, Germany \\
$^2$ CISPA Helmholtz Center for Information Security, Germany}

\begin{acronym}
\acro{aa}[AA]{adversarial attack}
\acro{db}[DB]{Diffusion Buffer}
\acro{rir}[RIR]{Room impulse response}
\acro{nn}[NN]{Neural Network}
\acro{flop}[FLOP]{Floating Point Operations}
\acro{dsm}[DSM]{Denoising Score Matching}
\acro{sgm}[SGM]{score-based generative model}
\acro{snr}[SNR]{signal-to-noise ratio}
\acro{gan}[GAN]{generative adversarial network}
\acro{vae}[VAE]{variational autoencoder}
\acro{ddpm}[DDPM]{denoising diffusion probabilistic model}
\acro{ncsnpp}[NCSN++]{Noise Conditional Score Network}
\acro{sncsnpp}[Sym-NCSN++]{Symmetric Noise Conditional Score Network}
\acro{stft}[STFT]{short-time Fourier transform}
\acro{istft}[iSTFT]{inverse short-time Fourier transform}
\acro{sde}[SDE]{stochastic differential equation}
\acro{ode}[ODE]{ordinary differential equation}
\acro{ou}[OU]{Ornstein-Uhlenbeck}
\acro{ve}[VE]{Variance Exploding}
\acro{dnn}[DNN]{deep neural network}
\acro{pesq}[PESQ]{Perceptual Evaluation of Speech Quality}
\acro{se}[SE]{Speech Enhancement}
\acro{bwe}[BWE]{bandwidth extension}
\acro{tf}[T-F]{time-frequency}
\acro{elbo}[ELBO]{evidence lower bound}
\acro{WPE}{weighted prediction error}
\acro{PSD}{power spectral density}
\acro{mse}[MSE]{Mean-Squared-Error}
\acro{POLQA}{Perceptual Objective Listening Quality Analysis}
\acro{SDR}{signal-to-distortion ratio}
\acro{estoi}[ESTOI]{Extended Short-Term Objective Intelligibility}
\acro{ph}[PH]{Psychoacoustic Hiding}
\acro{asr}[ASR]{automatic speech recognition}

\acro{wer}[WER]{Word Error Rate}
\acro{NFE}{number of function evaluations}
\acro{MOS}{mean opinion score}

\end{acronym}

\newcommand{\fSEd}{\ensuremath{f_{\mathrm{SE}\text{-}d}}\xspace}
\newcommand{\fSEM}{\ensuremath{f_{\mathrm{SE}\text{-}M}}\xspace}
\newcommand{\fSEdiff}{\ensuremath{f_{\mathrm{SE}\text{-}\mathrm{diff}}}\xspace}

\newcommand{\Suser}{\ensuremath{S^{\mathrm{user}}}}
\newcommand{\Sattacker}{\ensuremath{S^{\mathrm{attacker}}}}
\newcommand{\Yuser}{\ensuremath{Y^{\mathrm{user}}}}
\newcommand{\hatSattacked}{\ensuremath{\hat{S}^{Y^{\mathrm{user}}+\delta}}}

\newcommand{\HeatLight}{45}

\newcommand{\HeatCellUp}[4]{%
  \pgfmathsetmacro{\x}{#1}%
  \pgfmathsetmacro{\minv}{#2}%
  \pgfmathsetmacro{\maxv}{#3}%
  \pgfmathsetmacro{\pp}{100*(\x-\minv)/(\maxv-\minv)}%
  \pgfmathtruncatemacro{\HeatPct}{min(max(\pp,0),100)}%
  \edef\HeatSpec{green!\HeatPct!red!\HeatLight}%
  \expandafter\cellcolor\expandafter{\HeatSpec}#4%
}

\newcommand{\HeatCellDown}[4]{%
  \pgfmathsetmacro{\x}{#1}%
  \pgfmathsetmacro{\minv}{#2}%
  \pgfmathsetmacro{\maxv}{#3}%
  \pgfmathsetmacro{\pp}{100*(\maxv-\x)/(\maxv-\minv)}%
  \pgfmathtruncatemacro{\HeatPct}{min(max(\pp,0),100)}%
  \edef\HeatSpec{green!\HeatPct!red!\HeatLight}%
  \expandafter\cellcolor\expandafter{\HeatSpec}#4%
}

\begin{document}

\maketitle

\ninept

\begingroup
\renewcommand\thefootnote{}
\footnotetext{Project page (audio examples \& code): \url{https://sp-uhh.github.io/se-adversarial-attack}}
\endgroup
\begin{abstract}
Machine learning approaches for speech enhancement are becoming increasingly expressive, enabling ever more powerful modifications of input signals. In this paper, we demonstrate that this expressiveness introduces a vulnerability: advanced speech enhancement models can be susceptible to adversarial attacks. Specifically, we show that adversarial noise, carefully crafted and psychoacoustically masked by the original input, can be injected such that the enhanced speech output conveys an entirely different semantic meaning. We experimentally verify that contemporary predictive speech enhancement models can indeed be manipulated in this way. Furthermore, we highlight that diffusion models with stochastic samplers exhibit inherent robustness to such adversarial attacks by design.

\end{abstract}
\section{Introduction}
\label{sec:intro}

Adversarial attacks are a method where adversarial noise, designed to be hardly perceivable by humans, is added to the data such that the output of a \ac{dnn} is yielding a result that is controlled by the attacker but unintended by the user. A well-known example is the addition of adversarial noise to an image of a panda so that a classifier would detect a gibbon instead \cite{goodfellow2014explaining}. In speech processing, first attempts were made to attack \ac{asr} \cite{carlini2018audio} systems. For this, auditory masking effects can be exploited to psychoacoustically hide the adversarial noise in the speech commands \cite{schonherr2018adversarial}.

While adversarial attacks are most commonly designed for classification systems such as \ac{asr}, they can also be applied to regression tasks. %
For instance, one could think of a scenario where adversarial noise is added by an attacker, such that a \ac{se} system of a speech communication device such as a hearing aid or a telephony system outputs speech with a completely different and unintended semantic meaning.
However, modifying a speech signal such that the semantic meaning is deliberately changed by a \ac{se} system has received considerably less attention, also because traditional speech enhancement systems \cite{hendriks_dft-domain_2013}, e.g. based on Wiener filtering, are hardly expressive enough to change the semantic meaning of speech. However, now that both predictive and generative \ac{se} systems are becoming increasingly expressive, the risk of adversarial attacks increases. It is therefore important to investigate such potential vulnerabilities in more detail.

Therefore, in this paper, we explore the possibility of targeted adversarial attacks on modern expressive predictive and generative \ac{se} systems to produce an output with a completely different meaning than the original speech signal had. We consider a white-box scenario in which the adversary has full knowledge of the enhancement model and can compute gradients through it. 
We use a psychoacoustic model to make the adversarial noise as inaudible as possible, while keeping the intelligibility and quality of the enhanced speech as high as possible. We draw inspiration from psychoacoustic masking attacks in \ac{asr} \cite{schonherr2018adversarial}, extending them to the \ac{se} domain and addressing new challenges that arise, especially for diffusion-based generative models. The key contributions of this work are as follows:
(1) We propose an attack loss function and optimization procedure applicable to both traditional and diffusion-based \ac{se}, incorporating psychoacoustic constraints. 
(2) We analyze the differences in vulnerability between a standard predictive models like a regression \ac{se}, a mask-based \ac{se}, and a score-based diffusion \ac{se} (SGMSE+), including the effect of stochastic sampling in diffusion.
(3) We evaluate on EARS-WHAM-v2 using two metric groups: \emph{attack success}: DistillMOS and \ac{wer}, \ac{estoi}, \ac{POLQA} against the target speech; and \emph{perturbation impact}: \ac{POLQA}, \ac{estoi} between attacked and original mixtures, plus perturbation \ac{snr}.

This work raises an important and novel security issue by presenting a methodology for performing targeted adversarial examples against \ac{se} systems.

\section{Background}
\label{sec:se}
\acf{se} aims to recover a clean speech spectrogram $S$ from a noisy observation $Y$.
We work in the complex \ac{stft} domain and adopt the standard additive model
$$
    Y \;=\; S + N, \qquad S,\,Y,\,N \in \mathbb{C}^{F\times T},
$$
where $N$ denotes environmental noise. We consider two broad families of \ac{se} models. \emph{Predictive} approaches directly map $Y \mapsto \hat{S}$, either by regressing the complex spectrogram \cite{regression, fullyconvnns, ouyang2019fully} or by estimating a complex mask applied to $Y$ \cite{dcunet, crm}. \emph{Generative} approaches learn a prior over clean speech and perform conditional generation given $Y$; among these, diffusion (score-based) models have recently achieved state-of-the-art performance. For architectural consistency across comparisons, all our systems use the \ac{ncsnpp} backbone \cite{song2020score} adapted to complex inputs/outputs.

We introduce an \ac{se} model, denoted as $f_{SE}$, which serves as a generic operator that takes a noisy \ac{stft} $Y$ as input and aims to estimate the clean speech signal $S$, producing an output $\hat{S}^Y$.

\begin{equation}
    \hat{S}^Y = f_{SE}(Y)
\end{equation}

\subsection{Predictive models}
For training predictive models we use a point-wise complex-valued \ac{mse} loss to the clean target $S$, 
$$
    \mathcal{L}_{\text{reg}} \;=\; \sum_{q,n} \bigl|\hat{S}_{q,n}-S_{q,n}\bigr|^{2}.
$$
Here, $q$ and $n$ index frequency and time \ac{stft} bins respectively.
There are two common approaches to define a predictive model:
\paragraph*{Direct Mapping} is a common approach that directly maps the noisy spectrogram $Y$ to a clean estimate $\hat{S}$ with a \ac{nn} $d_{\theta}$ \cite{regression, fullyconvnns, ouyang2019fully}; we denote this by $\fSEd(Y)=d_{\theta}(Y)$.

\paragraph*{Masking (Complex Ratio Mask).}
In contrast, a mask network $M_{\theta}$ predicts a bounded complex ratio mask and applies it multiplicatively to the input; i.e., $\fSEM(Y)=M_{\theta}(Y)\odot Y$ (element-wise in the \ac{stft} domain) \cite{dcunet,crm}.
We use tanh-bounded logits to produce the mask: $M_r=2\tanh(\ell_{\Re})$, $M_i=2\tanh(\ell_{\Im})$, and $M=M_r+\mathrm{j}M_i$.

\subsection{Generative Speech Enhancement}
\label{sec:gse}

We adopt SGMSE+ \cite{richter_sgmse} for generative \ac{se}. It frames \ac{se} as a conditional generation task in the complex \ac{stft} domain. A continuous-time diffusion process is conditioned on the observation $Y$, and an \ac{ncsnpp} score network $s_\theta(x_t,y,t)$ learns to reverse the corruption. At test time, we obtain $\hat{S}$ by solving the reverse \ac{sde}
\begin{equation}
    d x_t = \Big[ -f(x_t,Y) + g(t)^2\,s_\theta(x_t,Y,t) \Big]dt + g(t)d\bar{w},
    \label{eq:rev-sde}
\end{equation}
where $f(x_t,Y)$ is the drift, $g(t)$ the diffusion coefficient and $\bar{w}$ denotes a standard Wiener process.

We define the diffusion \ac{se} model as \fSEdiff(Y), meaning that the enhanced estimate $\hat S$ is obtained by running a reverse \ac{sde} on \eqref{eq:rev-sde} from an initial state $x_T\!\sim\!\mathcal{N}_{\mathbb{C}}(Y,\sigma(T)^2 I)$ down to $t{=}0$ (conditioned on $Y$). In this notation, \fSEdiff\ is the integration operator over the entire reverse \ac{sde}.

In our experiments we report results under two sampling regimes: (1) the standard stochastic reverse \ac{sde}, where noise increments $d\bar{w}$ are drawn at each step; and (2) a \emph{fixed-noise} reverse \ac{sde}, in which a single Wiener path (random seed) is frozen for the duration of optimization. Freezing the path removes sampling variability and makes the trajectory deterministic, enabling controlled comparisons and stable gradient propagation in white-box adversarial attacks. This setup isolates the effect of diffusion stochasticity on robustness.

\section{Adversarial Attacks}
We consider targeted \emph{white-box} attacks: the attacker knows the enhancement system and can backpropagate through it down to the input. Given a source mixture $\Yuser=\Suser+N$ and a speech signal of comparable length targeted by the attacker $\Sattacker$ in the complex \ac{stft} domain, the goal is to add a small perturbation $\delta\in\mathbb{C}^{F\times T}$ to the source mixture $\Yuser$ so that the enhanced output $f_{SE}(\Yuser+\delta)$ resembles the clean speech $\Sattacker$ targeted by the attacker but unintended by the user.

Applying $f_{SE}$ in its direct, mask, or diffusion form: \fSEd, \fSEM, \fSEdiff to the adversarial mixture $\Yuser+\delta$ yields:
\begin{equation}
    \hatSattacked \;=\; f_{SE}\big(\Yuser + \delta\big).
    \label{eq:delta_inf}
\end{equation}

Then we optimize $\delta$ so that the SE system's output is close to $\Sattacker$ instead of \Suser using an \ac{mse} loss function 
\begin{equation}
    \mathcal{L}_{\mathrm{adv}}(\delta) \;=\; \sum_{q,n}\big| \hatSattacked_{q,n} - \Sattacker_{q,n}\big|^{2}.
    \label{eq:attack_objective}
\end{equation}

resulting in the iterative update 
\begin{equation}
    \delta^{(k+1)} \;\leftarrow\; \delta^{(k)} - \eta\,\nabla_{\delta^{(k)}}\,\mathcal{L}_{\mathrm{adv}}(\delta^{(k)}), \quad \quad \delta^{(0)}=0.
    \label{eq:delta_opt}
\end{equation}

Here, $\eta$ denotes the step size (learning rate), and $k=0,\dots,K\!-\!1$ indexes the iteration of the optimization. Furthermore, we employ gradient descent with momentum to stabilize and speed up the optimization.

\subsection{Backpropagation for diffusion models}
White-box attacks on diffusion models require gradients through every reverse step of \eqref{eq:rev-sde}. A straightforward approach saves activations and gradients at each step, so memory grows with the number of steps. We avoid this by using \emph{activation checkpointing}\footnote{\url{https://docs.pytorch.org/docs/stable/checkpoint.html}}: we skip saving intermediate states in the forward pass and recompute them during backpropagation, trading extra compute for lower memory. We checkpoint the score-network calls at each reverse step, which makes the memory requirement almost independent of the step count, at the cost of roughly $2\times$ more computation.

\subsection{Psychoacoustic Hiding}
\label{sec:ph}

Along with optimizing the adversarial loss \eqref{eq:attack_objective}, we adapt the \acf{ph} model proposed in \cite{schonherr2018adversarial}: the perturbation $\delta\!\in\!\mathbb{C}^{F\times T}$ is optimized using a psychoacoustic model, such that the added adversarial noise $\delta$ is masked by the mixture $\Yuser$.

\paragraph*{Hearing threshold $H$ (dB SPL).}
$H_{q,n}$ is the per-time-frequency audibility limit: it combines the absolute threshold in quiet with signal-dependent masking, so that perturbations kept below $H$ are hidden for the listener. Following \cite{schonherr2018adversarial}, we estimate $H$ from noisy observation $\Yuser$ with the MPEG-1 psychoacoustic model. For reproducibility, we use an open-source implementation\footnote{\url{https://github.com/RUB-SysSec/dompteur/tree/main/standalone-psychoacoustic-filtering}}.

\paragraph*{Spectral difference $D$ (dB).}
We define the perturbation per bin relative to the peak magnitude, as in \cite{schonherr2018adversarial}:
\begin{equation}
    D_{q,n} = 20\log_{10}\!\left(\frac{\big|\delta_{q,n}\big|}{\max_{q,n}(|\Yuser|)}\right).
\end{equation}
This yields a per-bin deviation in dB and we can directly compare it to $H_{q,n}$

\paragraph*{Mask $\Phi$.}
Following \cite{schonherr2018adversarial}, we quantify the level of distortion that is still acceptable $\Phi$ by comparing the per-bin spectral difference $D$ to the hearing threshold $H$:
\begin{equation}
    \Phi_{q,n} \;=\; H_{q,n} \;-\; D_{q,n}.
\end{equation}

Negative values for $\Phi_{q,n}$ indicate that the threshold is exceeded; positive values show the remaining headroom. We introduce a tolerance parameter $\lambda$ (in dB), and set negative values to zero:
\begin{equation}
    \Phi^{\!*}_{q,n} \;=\; \max\!\big(\Phi_{q,n} + \lambda,\, 0\big).
\end{equation}
The tolerance $\lambda$ allows to control the trade-off between imperceptibility and attack strength. Then, we obtain a mask $\hat{\Phi}\!\in\![0,1]$ by applying a standard min-max normalization.

\paragraph*{Gated optimization.}
Finally, we apply $\hat{\Phi}(q,n)$ as a psychoacoustic gate to the gradient:
\begin{align}
    \delta^{(k+1)} &\leftarrow \delta^{(k)} - \eta\,
    \Big(\hat{\Phi}^{(k)} \odot \nabla_{\delta^{(k)}}\,\mathcal{L}_{\mathrm{adv}}(\delta^{(k)})\Big)
\end{align}
Here, $\odot$ is element-wise multiplication. Because $\hat{\Phi}$ is real-valued, it scales the real and imaginary parts equally. In our implementation, the hearing threshold $H$ is computed once per utterance from $\Yuser$ and kept fixed during optimization; the spectral difference $D$ and the resulting masks $\hat{\Phi}$ are recomputed at every iteration as $\delta$ changes. %

\subsection{Projected Gradient Descent with an $\ell_2$ budget}
We found an $\ell_2$ constraint effective in practice. Accordingly, we optimize under a fixed $\ell_2$ budget using projected gradient descent (PGD)~\cite{madry2017towards}. We formulate the constrained optimization:
\begin{equation}
  \min_{\delta}\ \mathcal{L}_{\mathrm{adv}}(\delta)
  \quad\text{s.t.}\quad \|\delta\|_{2}\le \varepsilon,
  \label{eq:pgd_constraint}
\end{equation}
One PGD update is
\begin{equation}
  \delta^{(k+1)} \;=\;
  \Pi_{B_2(\varepsilon)}\!\Big(
     \delta^{(k)} - \eta\,\big(\hat{\Phi}^{(k)} \odot \nabla_{\delta^{(k)}}\mathcal{L}_{\mathrm{adv}}(\delta^{(k)})\big)
  \Big),
  \label{eq:pgd-one}
\end{equation}
where $\Pi_{{B}_2(\varepsilon)}$ is the projection onto the $\ell_2$ ball of radius $\varepsilon$. The corresponding projection results in:
$$
    \Pi_{B_2(\varepsilon)}(\delta)
    = \delta \cdot \min\!\Big\{1,\; \frac{\varepsilon}{\|\delta\|_2}\Big\},\qquad
    \|\delta\|_2=\sqrt{\sum_{q,n}\big|\delta_{q,n}\big|^2}.
$$

PGD solves the constrained attack by alternating a gradient step with a projection back onto the norm ball, such that in every iteration $\|\delta\|_2\le\varepsilon$ is satisfied. As a consequence, PGD limits the total energy of the attack, while $\hat{\Phi}$ spectrally shapes the attack such that it is masked by $\Yuser$.

\newcommand{\Best}[1]{\bm{#1}}

\newcommand{\DistillMOSMin}{1.68} \newcommand{\DistillMOSMax}{4.16}
\newcommand{\POLQAAttMin}{1.03}    \newcommand{\POLQAAttMax}{4.09}
\newcommand{\ESTOIAttMin}{0.11}    \newcommand{\ESTOIAttMax}{0.94}
\newcommand{\WERMin}{0.02}         \newcommand{\WERMax}{1.37}
\newcommand{\POLQAImpMin}{1.12}    \newcommand{\POLQAImpMax}{4.59}
\newcommand{\ESTOIImpMin}{0.14}    \newcommand{\ESTOIImpMax}{0.94}
\newcommand{\SNRMin}{-10.96}       \newcommand{\SNRMax}{32.46}

\newcommand{\DistillMOSCell}[2]{\HeatCellUp{#1}{\DistillMOSMin}{\DistillMOSMax}{#2}}
\newcommand{\POLQAAttCell}[2]{\HeatCellUp{#1}{\POLQAAttMin}{\POLQAAttMax}{#2}}
\newcommand{\ESTOIAttCell}[2]{\HeatCellUp{#1}{\ESTOIAttMin}{\ESTOIAttMax}{#2}}
\newcommand{\WERCell}[2]{\HeatCellDown{#1}{\WERMin}{\WERMax}{#2}}
\newcommand{\POLQAImpCell}[2]{\HeatCellUp{#1}{\POLQAImpMin}{\POLQAImpMax}{#2}}
\newcommand{\ESTOIImpCell}[2]{\HeatCellUp{#1}{\ESTOIImpMin}{\ESTOIImpMax}{#2}}
\newcommand{\SNRCell}[2]{\HeatCellUp{#1}{\SNRMin}{\SNRMax}{#2}}

\begin{table*}[t]
\centering
\scriptsize
\setlength{\tabcolsep}{4.5pt}
\renewcommand{\arraystretch}{1.05}
\begin{tabular}{l l c c
                c c c c
                c c c}
\toprule
& & & &
\multicolumn{4}{c}{\textbf{Attack success}} &
\multicolumn{3}{c}{\textbf{Perturbation impact}} \\
\cmidrule(lr){5-8}\cmidrule(lr){9-11}
\makecell{Family} & \makecell{Model} & $\lambda$ & $\varepsilon$ &
\makecell{DistillMOS $\uparrow$} &
\makecell{POLQA $\uparrow$} &
\makecell{ESTOI $\uparrow$} &
\makecell{WER $\downarrow$} &
\makecell{POLQA $\uparrow$} &
\makecell{ESTOI $\uparrow$} &
\makecell{SNR (dB) $\uparrow$} \\
\cmidrule(lr){6-8}\cmidrule(lr){9-10}
& & & &
\makecell{\scriptsize $\hatSattacked$} &
\multicolumn{3}{c}{\scriptsize $\hatSattacked$ vs.\ $\Sattacker$} &
\multicolumn{2}{c}{\scriptsize ($\Yuser{+}\delta$) vs.\ $\Yuser$} &
\makecell{\scriptsize $\Yuser$/$\delta$} \\
\midrule

\multirow{13}{*}{\rotatebox{90}{\textbf{Predictive}}}
& \multirow{13}{*}{\textbf{Direct Map}}
& —   &  $\infty$   &
\DistillMOSCell{4.16}{$4.16 \pm 0.56$} &
\POLQAAttCell{4.09}{$4.09 \pm 0.61$} &
\ESTOIAttCell{0.94}{$0.94 \pm 0.09$} &
\WERCell{0.02}{$0.02 \pm 0.04$} &
\POLQAImpCell{1.34}{$1.34 \pm 0.40$} &
\ESTOIImpCell{0.25}{$0.25 \pm 0.08$}  &
\SNRCell{-2.89}{$-2.89 \pm 5.49$} \\
\cmidrule(lr){3-11}
& & —   & 10  &
\DistillMOSCell{3.42}{$\Best{3.42 \pm 0.62}$} &
\POLQAAttCell{2.99}{$\Best{2.99 \pm 0.70}$} &
\ESTOIAttCell{0.81}{$\Best{0.81 \pm 0.13}$} &
\WERCell{0.15}{$\Best{0.15 \pm 0.90}$} &
\POLQAImpCell{2.72}{$2.72 \pm 0.81$} &
\ESTOIImpCell{0.65}{$0.65 \pm 0.12$} &
\SNRCell{8.18}{$8.18 \pm 5.80$} \\
& & 40  & 10  &
\DistillMOSCell{3.27}{$3.27 \pm 0.67$} &
\POLQAAttCell{2.62}{$2.62 \pm 0.73$} &
\ESTOIAttCell{0.77}{$0.77 \pm 0.16$} &
\WERCell{0.20}{$0.20 \pm 0.90$} &
\POLQAImpCell{2.98}{$2.98 \pm 0.79$} &
\ESTOIImpCell{0.67}{$0.67 \pm 0.13$} &
\SNRCell{10.03}{$10.03 \pm 4.99$} \\
& & 20  & 10  &
\DistillMOSCell{2.54}{$2.54 \pm 0.51$} &
\POLQAAttCell{1.81}{$1.81 \pm 0.43$} &
\ESTOIAttCell{0.68}{$0.68 \pm 0.11$} &
\WERCell{0.20}{$0.20 \pm 1.02$} &
\POLQAImpCell{3.19}{$\Best{3.19 \pm 0.85}$} &
\ESTOIImpCell{0.70}{$0.70 \pm 0.12$} &
\SNRCell{12.88}{$\Best{12.88 \pm 5.13}$} \\
& & 10  & 10  &
\DistillMOSCell{1.98}{$1.98 \pm 0.29$} &
\POLQAAttCell{1.22}{$1.22 \pm 0.14$} &
\ESTOIAttCell{0.48}{$0.48 \pm 0.10$} &
\WERCell{0.49}{$0.49 \pm 1.32$} &
\POLQAImpCell{3.05}{$3.05 \pm 0.99$} &
\ESTOIImpCell{0.72}{$\Best{0.72 \pm 0.12}$} &
\SNRCell{12.05}{$12.05 \pm 6.63$} \\
& & 0   & 10  &
\DistillMOSCell{1.72}{$1.72 \pm 0.20$} &
\POLQAAttCell{1.07}{$1.07 \pm 0.04$} &
\ESTOIAttCell{0.22}{$0.22 \pm 0.07$} &
\WERCell{1.37}{$1.37 \pm 1.89$} &
\POLQAImpCell{2.55}{$2.55 \pm 0.93$} &
\ESTOIImpCell{0.69}{$0.69 \pm 0.12$} &
\SNRCell{7.92}{$7.92 \pm 5.90$} \\
\cmidrule(lr){3-11}
& & 20  &  $\infty$   &
\DistillMOSCell{2.36}{$2.36 \pm 0.46$} &
\POLQAAttCell{1.64}{$1.64 \pm 0.36$} &
\ESTOIAttCell{0.65}{$0.65 \pm 0.12$} &
\WERCell{0.23}{$0.23 \pm 1.06$} &
\POLQAImpCell{2.40}{$2.40 \pm 0.60$} &
\ESTOIImpCell{0.58}{$0.58 \pm 0.09$} &
\SNRCell{7.72}{$7.72 \pm 5.38$} \\
& & 20  & 20  &
\DistillMOSCell{2.87}{$\Best{2.87 \pm 0.64}$} &
\POLQAAttCell{1.94}{$1.94 \pm 0.59$} &
\ESTOIAttCell{0.71}{$0.71 \pm 0.12$} &
\WERCell{0.19}{$0.19 \pm 0.89$} &
\POLQAImpCell{1.84}{$1.84 \pm 0.78$} &
\ESTOIImpCell{0.40}{$0.40 \pm 0.15$} &
\SNRCell{0.50}{$0.50 \pm 6.56$} \\
& & 20  & 15  &
\DistillMOSCell{2.86}{$2.86 \pm 0.57$} &
\POLQAAttCell{2.00}{$\Best{2.00 \pm 0.52}$} &
\ESTOIAttCell{0.73}{$\Best{0.73 \pm 0.10}$} &
\WERCell{0.17}{$\Best{0.17 \pm 0.90}$} &
\POLQAImpCell{2.26}{$2.26 \pm 0.93$} &
\ESTOIImpCell{0.52}{$0.52 \pm 0.15$} &
\SNRCell{5.64}{$5.64 \pm 5.98$} \\
& & 20  & 10  &
\DistillMOSCell{2.54}{$2.54 \pm 0.51$} &
\POLQAAttCell{1.81}{$1.81 \pm 0.43$} &
\ESTOIAttCell{0.68}{$0.68 \pm 0.11$} &
\WERCell{0.20}{$0.20 \pm 1.02$} &
\POLQAImpCell{3.19}{$3.19 \pm 0.85$} &
\ESTOIImpCell{0.70}{$0.70 \pm 0.12$} &
\SNRCell{12.88}{$12.88 \pm 5.13$} \\
& & 20  & 6   &
\DistillMOSCell{1.95}{$1.95 \pm 0.32$} &
\POLQAAttCell{1.30}{$1.30 \pm 0.18$} &
\ESTOIAttCell{0.50}{$0.50 \pm 0.13$} &
\WERCell{0.45}{$0.45 \pm 1.05$} &
\POLQAImpCell{4.03}{$4.03 \pm 0.61$} &
\ESTOIImpCell{0.84}{$0.84 \pm 0.08$} &
\SNRCell{21.18}{$21.18 \pm 5.05$} \\
& & 20  & 3   &
\DistillMOSCell{1.68}{$1.68 \pm 0.24$} &
\POLQAAttCell{1.09}{$1.09 \pm 0.07$} &
\ESTOIAttCell{0.26}{$0.26 \pm 0.13$} &
\WERCell{1.17}{$1.17 \pm 0.85$} &
\POLQAImpCell{4.59}{$\Best{4.59 \pm 0.33}$} &
\ESTOIImpCell{0.94}{$\Best{0.94 \pm 0.04}$} &
\SNRCell{32.46}{$\Best{32.46 \pm 5.34}$} \\
\cmidrule(lr){2-11}

& \multirow{3}{*}{\textbf{CRM}}
& 40 & 10 &
\DistillMOSCell{2.92}{$\Best{2.92 \pm 0.60}$} &
\POLQAAttCell{2.32}{$\Best{2.32 \pm 0.69}$} &
\ESTOIAttCell{0.73}{$\Best{0.73 \pm 0.16}$} &
\WERCell{0.23}{$\Best{0.23 \pm 0.92}$} &
\POLQAImpCell{2.96}{$2.96 \pm 0.78$} &
\ESTOIImpCell{0.67}{$0.67 \pm 0.13$} &
\SNRCell{10.43}{$10.43 \pm 4.84$} \\
& & 20 & 10 &
\DistillMOSCell{2.19}{$2.19 \pm 0.41$} &
\POLQAAttCell{1.57}{$1.57 \pm 0.32$} &
\ESTOIAttCell{0.64}{$0.64 \pm 0.10$} &
\WERCell{0.23}{$\Best{0.23 \pm 0.90}$} &
\POLQAImpCell{3.14}{$\Best{3.14 \pm 0.84}$} &
\ESTOIImpCell{0.69}{$\Best{0.69 \pm 0.12}$} &
\SNRCell{12.90}{$\Best{12.90 \pm 5.35}$} \\
& & 0  & 10 &
\DistillMOSCell{1.70}{$1.70 \pm 0.21$} &
\POLQAAttCell{1.07}{$1.07 \pm 0.03$} &
\ESTOIAttCell{0.20}{$0.20 \pm 0.07$} &
\WERCell{1.34}{$1.34 \pm 1.31$} &
\POLQAImpCell{2.58}{$2.58 \pm 0.91$} &
\ESTOIImpCell{0.69}{$\Best{0.69 \pm 0.11}$} &
\SNRCell{8.53}{$8.53 \pm 6.13$} \\
\midrule

\multirow{4}{*}{\rotatebox{90}{\textbf{Generative}}}
& \multirow{4}{*}{\makecell{\textbf{Diffusion}}}
& —  & $\infty$ &
\DistillMOSCell{3.40}{${3.40 \pm 1.06}$} &
\POLQAAttCell{2.28}{${2.28 \pm 1.19}$} &
\ESTOIAttCell{0.69}{${0.69 \pm 0.26}$} &
\WERCell{0.47}{${0.47 \pm 1.01}$} &
\POLQAImpCell{1.12}{$1.12 \pm 0.14$} &
\ESTOIImpCell{0.14}{$0.14 \pm 0.07$} &
\SNRCell{-10.96}{$-10.96 \pm 6.79$} \\
\cmidrule(lr){3-11}
& & 40 & 10 &
\DistillMOSCell{2.86}{$\Best{2.86 \pm 0.94}$} &
\POLQAAttCell{1.45}{$\Best{1.45 \pm 0.59}$} &
\ESTOIAttCell{0.34}{$\Best{0.34 \pm 0.29}$} &
\WERCell{0.71}{$\Best{0.71 \pm 0.48}$} &
\POLQAImpCell{2.72}{$2.72 \pm 0.84$} &
\ESTOIImpCell{0.68}{$0.68 \pm 0.14$} &
\SNRCell{6.59}{$6.59 \pm 4.99$} \\
& & 20 & 10 &
\DistillMOSCell{2.12}{$2.12 \pm 0.63$} &
\POLQAAttCell{1.14}{$1.14 \pm 0.19$} &
\ESTOIAttCell{0.24}{$0.24 \pm 0.22$} &
\WERCell{0.80}{$0.80 \pm 0.45$} &
\POLQAImpCell{2.80}{$2.80 \pm 0.78$} &
\ESTOIImpCell{0.70}{$0.70 \pm 0.11$} &
\SNRCell{7.90}{$7.90 \pm 5.96$} \\
& & 10 & 10 &
\DistillMOSCell{1.84}{$1.84 \pm 0.43$} &
\POLQAAttCell{1.03}{$1.03 \pm 0.04$} &
\ESTOIAttCell{0.11}{$0.11 \pm 0.11$} &
\WERCell{1.15}{$1.15 \pm 0.36$} &
\POLQAImpCell{3.17}{$\Best{3.17 \pm 0.64}$} &
\ESTOIImpCell{0.77}{$\Best{0.77 \pm 0.06}$} &
\SNRCell{10.26}{$\Best{10.26 \pm 5.32}$} \\
\bottomrule
\end{tabular}

\caption{Targeted-attack results on 100 EARS-WHAM-v2 pairs. Left block of metrics reports attack success for the enhanced signal $\hatSattacked$ against the target $\Sattacker$; right block reports perturbation impact for $(\Yuser{+}\delta)$ against $\Yuser$ and the SNR of $\delta$ w.r.t.\ $\Yuser$ (higher SNR = quieter perturbation). Cell backgrounds show a per-metric min-max colormap (respecting $\uparrow/\downarrow$), and the best value within each block is bolded. Values are reported as mean $\pm$ standard deviation.}

\label{tab:results-attack-impact-compact}
\end{table*}

\section{Experimental setup}

\subsection{Model preparation}
We adopt the SGMSE$+$ pipeline\footnote{\url{https://github.com/sp-uhh/sgmse}} as our baseline repository to obtain a strong diffusion-based \ac{se} system. We prepared a generative diffusion model and two predictive baselines that reuse the same backbone but differ only in how $\hat S$ is produced (Section~\ref{sec:se}). We took the \ac{ncsnpp} U-Net backbone as a baseline architecture \cite{song2020score}. To reduce the computational complexity, we modified the network depth by decreasing the number of residual blocks from 2 to 1. Unless noted otherwise, the remaining SGMSE$+$ settings are left unchanged.

\textbf{Model variants.}
\begin{itemize}
    \item \textbf{Direct Map} - a predictive network $d_{\theta}$ that maps $Y$ to $\hat S$, trained with the point-wise complex \ac{mse} to the clean target $S$.
    \item \textbf{Complex Ratio Mask} - a predictive mask model $M_{\theta}$ that predicts a bounded complex mask and applies it to $Y$, trained with the same complex \ac{mse}.
    \item \textbf{Diffusion} - a score-based SGMSE$+$ model with the OUVE configuration \cite{richter_sgmse}. At inference we run the reverse \ac{sde} sampler, i.e., $\fSEdiff(Y)$.
\end{itemize}

Training and evaluation are conducted on EARS–WHAM-v2 dataset \cite{richter2024ears} with 86 hours of clean and noisy speech in the training part. In this corpus, mixtures are defined at SNRs uniformly sampled from $[-2.5, 17.5]$ dB.

\subsection{Adversarial Attack with Psychoacoustic Hiding}

From the EARS-WHAM-v2 test set (850 utterances; 6 speakers, 3 male / 3 female) we sample $100$ random pairs $(\Yuser, \Sattacker)$ of comparable duration, where $\Yuser$ is a noisy mixture from utterance~1 and $\Sattacker$ is a clean utterance~2. For each pair, we optimize a complex-valued perturbation $\delta \in \mathbb{C}^{F\times T}$ and construct the attacked input $\Yuser+\delta$ and output $\hatSattacked$ \eqref{eq:delta_inf}. Starting with $\delta^{(0)}\!=\!0$, we update $\delta$ with the PGD rule \eqref{eq:pgd-one}, using $K{=}150$ iterations of SGD with momentum $0.4$ and learning rate $0.1$, while keeping model parameters fixed. 
For the diffusion-based \ac{se}, we set the number of reverse steps to $N{=}25$.

Imperceptibility is promoted via the psychoacoustic gate from Sec.~\ref{sec:ph}, computed from $\Yuser$. A tolerance parameter $\lambda \in \{0,\allowbreak 10,\allowbreak 20,\allowbreak 40\}\,\mathrm{dB}$ shifts the masking threshold - larger $\lambda$ relaxes the constraint.  We additionally impose an $\ell_2$ budget on $\delta$ via the projection radius $\varepsilon$.

\subsection{Evaluation metrics}
We evaluate adversarial attacks along two axes: \emph{attack success} - how well the output signal $\hatSattacked$ matches the target speech $\Sattacker$; and \emph{perturbation impact} - how audible the perturbation $\delta$ is within the original mixture $\Yuser$.

\textbf{Attack success (AS).}
We report \ac{wer}, \ac{POLQA}, \ac{estoi} between $\hatSattacked$ and $\Sattacker$, and DistillMOS~\cite{stahl2025distillation} for $\hatSattacked$. For WER, we transcribe with five Whisper~\cite{radford2023robust} variants (Large, Large-v1, Large-v2, Large-v3, Large-v3-turbo) and average the resulting WERs for stability.

\textbf{Perturbation impact (PI).}
To quantify how much the input mixture is altered, we use \ac{POLQA} and \ac{estoi} as a distance metric between $(\Yuser+\delta)$ and $\Yuser$. 

In addition, we report the signal-to-noise ratio (SNR, dB) of $\Yuser$ relative to $\delta$~\cite{carlini2018audio}:
$$
    \mathrm{SNR} \;=\; 20\log_{10}\!\frac{\|\Yuser\|_2}{\|\delta\|_2}.
$$
Here $\|\cdot\|_2$ denotes the Euclidean norm. A higher SNR indicates that the adversarial attack is of lower power.

\newcommand{\DistillMOSAbCell}[2]{\HeatCellUp{#1}{\DistillMOSMin}{\DistillMOSMax}{#2}}
\newcommand{\POLQAAbCell}[2]{\HeatCellUp{#1}{\POLQAAttMin}{\POLQAAttMax}{#2}}
\newcommand{\ESTOIAbCell}[2]{\HeatCellUp{#1}{\ESTOIAttMin}{\ESTOIAttMax}{#2}}
\newcommand{\WERAbCell}[2]{\HeatCellDown{#1}{\WERMin}{\WERMax}{#2}}
\newcommand{\SNRAbCell}[2]{\HeatCellUp{#1}{\SNRMin}{\SNRMax}{#2}}

\begin{table}[t]
\centering
\scriptsize
\renewcommand{\arraystretch}{1.05}
\setlength{\tabcolsep}{3.2pt}
\begin{tabular}{@{} l c c c c c @{}}
\toprule
& \multicolumn{4}{c}{\textbf{Attack success}} &
\multicolumn{1}{c}{\textbf{Perturbation impact}} \\
\cmidrule(lr){2-5}\cmidrule(lr){6-6}
\makecell{Variant} &
\makecell{DistillMOS $\uparrow$} &
\makecell{POLQA $\uparrow$} &
\makecell{ESTOI $\uparrow$} &
\makecell{WER $\downarrow$} &
\makecell{SNR (dB) $\uparrow$} \\
\midrule
Stochastic &
\DistillMOSAbCell{3.40}{$3.40$} &
\POLQAAbCell{2.28}{$2.28$} &
\ESTOIAbCell{0.69}{$0.69$} &
\WERAbCell{0.47}{$0.47$} &
\SNRAbCell{-10.96}{$-10.96$} \\
Fixed-noise &
\DistillMOSAbCell{3.90}{$\Best{3.90}$} &
\POLQAAbCell{3.03}{$\Best{3.03}$} &
\ESTOIAbCell{0.81}{$0.81$} &
\WERAbCell{0.27}{$0.27$} &
\SNRAbCell{-7.73}{$-7.73$} \\
$N{=}15$ &
\DistillMOSAbCell{3.69}{$3.69$} &
\POLQAAbCell{2.99}{$2.99$} &
\ESTOIAbCell{0.82}{$\Best{0.82}$} &
\WERAbCell{0.22}{$\Best{0.22}$} &
\SNRAbCell{-6.61}{$\Best{-6.61}$} \\
$N{=}35$ &
\DistillMOSAbCell{3.13}{$3.13$} &
\POLQAAbCell{1.92}{$1.92$} &
\ESTOIAbCell{0.61}{$0.61$} &
\WERAbCell{0.57}{$0.57$} &
\SNRAbCell{-13.46}{$-13.46$} \\
$\sigma_{\mathrm{max}}{=}0.3$ &
\DistillMOSAbCell{2.72}{$2.72$} &
\POLQAAbCell{1.61}{$1.61$} &
\ESTOIAbCell{0.50}{$0.50$} &
\WERAbCell{0.69}{$0.69$} &
\SNRAbCell{-15.33}{$-15.33$} \\
$\sigma_{\mathrm{max}}{=}0.7$ &
\DistillMOSAbCell{3.63}{$3.63$} &
\POLQAAbCell{2.77}{$2.77$} &
\ESTOIAbCell{0.78}{$0.78$} &
\WERAbCell{0.28}{$0.28$} &
\SNRAbCell{-11.51}{$-11.51$} \\
\bottomrule
\end{tabular}
\caption{Ablation of an unconstrained adversarial attack on diffusion \ac{se} ($\lambda{=}-, \varepsilon{=}\infty$). Unless specified, the sampler uses $N{=}25$ reverse steps and stochastic sampling; \emph{Fixed-noise} denotes a frozen noise path (Sec.~\ref{sec:gse}). The baseline model is trained with $\sigma_{\mathrm{max}}{=}0.5$; rows that list $\sigma_{\mathrm{max}}$ are separately trained models. Cell backgrounds show a min--max colormap per metric using the same color ranges as Table~\ref{tab:results-attack-impact-compact}}
\label{tab:ablation-ouve}
\end{table}

\section{Results and Analysis}
Table~\ref{tab:results-attack-impact-compact} summarizes targeted attacks on 100 EARS-WHAM-v2 pairs. Below we highlight the main results.

\textbf{Direct Map (predictive).}
Unconstrained attacks ($\lambda{=}-, \varepsilon{=}\infty$) steer almost perfectly (e.g., WER$\approx$0.02, AS-ESTOI$ \approx$0.94) while being clearly audible SNR$\approx$-2.9\ dB. Introducing constraints exposes a clean two–knob trade-off:

\emph{(1) Fixed energy ($\varepsilon{=}10$), sweep $\lambda$:} with the energy budget fixed, SNR and perturbation impact metrics stay within a narrow band (SNR $\approx 8-13\ {dB}$; PI-POLQA $\approx 2.55-3.19$; PI-ESTOI $\approx 0.65-0.72$), while \emph{attack success} shifts significantly - e.g., WER $\approx 0.20 \to 1.37$, AS–ESTOI $\approx 0.77 \to 0.22$, AS–POLQA $\approx 2.62 \to 1.07$ (DistillMOS $\approx 3.27 \to 1.72$) as $\lambda$ tightens ($40 \to 0$).

\emph{(2) Fixed masking ($\lambda{=}20$), sweep $\varepsilon$:} enlarging $\varepsilon$ improves alignment (WER $1.17\to0.23$ from $\varepsilon{=}3\to\infty$) while increasing audibility (PI-POLQA $4.59\to2.40$; SNR $32.46\to7.72$ dB). Practically, moderate settings (e.g., $\lambda{=}20$, $\varepsilon\in\{10,15\}$) sit in a balanced region (WER $\approx 0.20-0.17$, PI-POLQA $\approx 3.19-2.26$, SNR $\approx 12.88-5.64$ dB).

\textbf{CRM (predictive, mask).}
For identical $(\lambda,\varepsilon)$, \emph{perturbation impact} matches Direct Map closely (e.g., at $\lambda{=}20,\varepsilon{=}10$: PI-POLQA $3.14$ vs.\ $3.19$; SNR $12.90$ vs.\ $12.88$ dB). \emph{Attack success} is consistently, but only slightly lower (e.g., DistillMOS $2.54$ vs.\ $2.19$, AS-POLQA $1.81$ vs.\ $1.57$, WER $0.23$ vs.\ $0.20$ at the same setting). In short, the masking approach with similar perturbation impact metrics produces a worse attack.

\textbf{Diffusion (generative).}
Diffusion is harder to control and hides worse under comparable or even looser constraints. Without constraints it injects more energy yet underperforms (WER$\approx$0.47; SNR$\approx$${-}10.96$ dB). With the same energy budget as in Direct Map ($\varepsilon{=}10$), it remains worse on both axes: attack audibility is higher (at $\lambda{=}20$: SNR $7.90$ vs.\ $12.88$ dB; PI-POLQA $2.80$ vs.\ $3.19$), and target match is weaker (AS-POLQA $1.14$ vs.\ $1.81$, WER $0.80$ vs.\ $0.20$).

\textbf{Diffusion ablations.} We conducted a small ablation study on diffusion \ac{se} under the unconstrained attack setting to examine how the diffusion model parameters affect attack success in the absence of explicit constraints (Table~\ref{tab:ablation-ouve}).
Fixing the noise path makes attacks more effective (WER $0.47\!\to\!0.27$) by removing protective randomness. Fewer reverse steps ease steering (e.g., $N{=}15$: WER$\approx$0.22) while more steps harden it ($N{=}35$: WER$\approx$0.57). Increasing $\sigma_{\mathrm{max}}$ facilitates attacks ($0.7$: WER$\approx$0.28) whereas decreasing it suppresses them ($0.3$: WER$\approx$0.69). The robustness pattern generalizes beyond WER and is consistently observed across all five metrics.

\section{Conclusions}
In this paper, we show that modern speech enhancement systems exhibit some vulnerability to adversarial attacks. That is, by adding adversarial noise, an attacker can trick a speech enhancement system to output a completely different signal than intended by the user. To that end, we build white-box attacks with psychoacoustic masking and an $\ell_2$ budget for the adversarial attack. We test predictive (direct map, complex mask) and diffusion SE on EARS-WHAM-v2 using success and audibility metrics.

We show that predictive SE is comparably easily attacked with a relatively small $\ell_2$ budget for the adversarial noise. The complex mask variant behaves similarly to direct mapping but is somewhat less vulnerable. Diffusion SE is more resistant to targeted manipulation: even with a larger $\ell_2$ budget for the adversarial noise, the output aligns less well with the attacker's target signal and remains more audible than predictive baselines. We show that stochastic sampling in the reverse diffusion process increases robustness to attacks even further.

\vfill\pagebreak

\section{Acknowledgements}
Funded by the Deutsche Forschungsgemeinschaft (DFG, German Research Foundation) -- 545210893, 498394658. The authors gratefully acknowledge the scientific support and HPC resources provided by the Erlangen National High Performance Computing Center (NHR@FAU) of the Friedrich-Alexander-Universität Erlangen-Nürnberg (FAU) under the NHR project f102ac. NHR funding is provided by federal and Bavarian state authorities. NHR@FAU hardware is partially funded by the German Research Foundation (DFG) – 440719683. This research was partially funded by VolkswagenStiftung Niedersächsisches Vorab - ZN4704 and the Daimler and Benz Foundation under the grant \emph{Ladenburger Kolleg, Project KonCheck}. We acknowledge funding by the German Federal Ministry of Research, Technology and Space (BMFTR) under grant agreement No. 16IS24072B (COMFORT), SisWiss (16KIS2330), and AIgenCY (16KIS2012). The authors would like to thank J. Berger and Rohde\&Schwarz SwissQual AG for their support with POLQA.

\bibliographystyle{IEEEbib}
\bibliography{strings,refs}

\begin{thebibliography}{10}

\bibitem{goodfellow2014explaining}
Ian~J Goodfellow, Jonathon Shlens, and Christian Szegedy,
\newblock ``Explaining and harnessing adversarial examples,''
\newblock in {\em Int. Conf. on Learning Representations (ICLR)}, San Diego,
  CA, USA, 2015.

\bibitem{carlini2018audio}
Nicholas Carlini and David Wagner,
\newblock ``Audio adversarial examples: Targeted attacks on speech-to-text,''
\newblock in {\em IEEE security and privacy workshops (SPW)}, San Francisco,
  CA, USA, 2018.

\bibitem{schonherr2018adversarial}
Lea Sch\"{o}nherr, Katharina Kohls, Steffen Zeiler, Thorsten Holz, and Dorothea
  Kolossa,
\newblock ``Adversarial attacks against automatic speech recognition systems
  via psychoacoustic hiding,''
\newblock in {\em Network and Distributed System Security Symposium (NDSS)},
  San Diego, CA, USA, 2019.

\bibitem{hendriks_dft-domain_2013}
Richard~C. Hendriks, Timo Gerkmann, and Jesper Jensen,
\newblock {\em {DFT}-Domain Based Single-icrophone Noise Reduction for Speech
  Enhancement: A Survey of the State-of-the-Art},
\newblock Number~11 in Synthesis Lectures on Speech and Audio Processing.
  {Morgan \& Claypool}, {Williston, VT}, 2013.

\bibitem{regression}
Yong Xu, Jun Du, Li-Rong Dai, and Chin-Hui Lee,
\newblock ``A regression approach to speech enhancement based on deep neural
  networks,''
\newblock {\em IEEE/ACM Trans. on Audio, Speech, and Language Proc. (TASLP)},
  vol. 23, no. 1, pp. 7--19, 2015.

\bibitem{fullyconvnns}
Se~Rim Park and Jinwon Lee,
\newblock ``A fully convolutional neural network for speech enhancement,''
\newblock in {\em ISCA Interspeech}, Stockholm, Sweden, 2017.

\bibitem{ouyang2019fully}
Zhiheng Ouyang, Hongjiang Yu, Wei-Ping Zhu, and Benoit Champagne,
\newblock ``A fully convolutional neural network for complex spectrogram
  processing in speech enhancement,''
\newblock in {\em IEEE Int. Conf. on Acoustics, Speech and Signal Proc.
  (ICASSP)}, Brighton, UK, 2019.

\bibitem{dcunet}
Hyeong-Seok Choi, Jang-Hyun Kim, Jaesung Huh, Adrian Kim, Jung-Woo Ha, and
  Kyogu Lee,
\newblock ``Phase-aware speech enhancement with deep complex {U-Net},''
\newblock in {\em Int. Conf. on Learning Representations (ICLR)}, Vancouver,
  Canada, 2018.

\bibitem{crm}
Donald~S. Williamson, Yuxuan Wang, and DeLiang Wang,
\newblock ``Complex ratio masking for monaural speech separation,''
\newblock {\em IEEE/ACM Trans. on Audio, Speech, and Language Proc. (TASLP)},
  vol. 24, no. 3, pp. 483--492, 2016.

\bibitem{song2020score}
Yang Song, Jascha Sohl-Dickstein, Diederik~P Kingma, Abhishek Kumar, Stefano
  Ermon, and Ben Poole,
\newblock ``Score-based generative modeling through stochastic differential
  equations,''
\newblock in {\em Int. Conf. on Learning Representations (ICLR)}, Vienna,
  Austria, 2021.

\bibitem{richter_sgmse}
Julius Richter, Simon Welker, Jean-Marie Lemercier, Bunlong Lay, and Timo
  Gerkmann,
\newblock ``Speech enhancement and dereverberation with diffusion-based
  generative models,''
\newblock {\em IEEE/ACM Trans. on Audio, Speech, and Language Proc. (TASLP)},
  vol. 31, pp. 2351–2364, 2023.

\bibitem{madry2017towards}
Aleksander Madry, Aleksandar Makelov, Ludwig Schmidt, Dimitris Tsipras, and
  Adrian Vladu,
\newblock ``Towards deep learning models resistant to adversarial attacks,''
\newblock in {\em Int. Conf. on Learning Representations (ICLR)}, Vancouver,
  Canada, 2018.

\bibitem{richter2024ears}
Julius Richter, Yi-Chiao Wu, Steven Krenn, Simon Welker, Bunlong Lay, Shinjii
  Watanabe, Alexander Richard, and Timo Gerkmann,
\newblock ``{EARS}: An anechoic fullband speech dataset benchmarked for speech
  enhancement and dereverberation,''
\newblock in {\em ISCA Interspeech}, Kos Island, Greece, 2024.

\bibitem{stahl2025distillation}
Benjamin Stahl and Hannes Gamper,
\newblock ``Distillation and pruning for scalable self-supervised
  representation-based speech quality assessment,''
\newblock in {\em IEEE Int. Conf. on Acoustics, Speech and Signal Proc.
  (ICASSP)}, Hyderabad, India, 2025.

\bibitem{radford2023robust}
Alec Radford, Jong~Wook Kim, Tao Xu, Greg Brockman, Christine McLeavey, and
  Ilya Sutskever,
\newblock ``Robust speech recognition via large-scale weak supervision,''
\newblock in {\em Int. Conf. on Learning Representations (ICLR)}, Kigali,
  Rwanda, 2023, PMLR.

\end{thebibliography}

\end{document}